\documentstyle[preprint,epsf,aps]{revtex}
\begin{document}
\title{ A remark on quantum key distribution with
two way communication: the classical complexity in decoding the CSS code can be removed
}
\author{Wang Xiang-bin\thanks{email: wang$@$qci.jst.go.jp} 
\\
        Imai Quantum Computation and Information project, ERATO, Japan Sci. and Tech. Corp.\\
Daini Hongo White Bldg. 201, 5-28-3, Hongo, Bunkyo, Tokyo 113-0033, Japan}

\maketitle 
\begin{abstract}
So far all the proven unconditionally secure
 prepare and measure protocols for the quantum key distribution(QKD) 
must solve the
very complex problem of decoding the classical CSS code. In the decoding stage,
Bob has to compare his string  with an exponentially large number of
all the strings in certain code space to find out the closest one.
 Here we have spotted that, 
in an entanglement purification
protocol(EPP), the random basis in the state preparation stage is only 
necessary to those check qubits, but uncessary to the code qubits. 
In our modified two way communication EPP(2-EPP) protocol,
Alice and Bob may first take all the  parity checks on $Z$ basis
to reduce the bit flip error to strictly zero with a high probability, e.g.,
$1-2^{-30}$, and then use the CSS code to obtain the final key. We show that,
this type  of 2-EPP protocol can be reduced to an equivalent prepare and measure protocol. In our protocol, the huge  complexity of decoding the classical
CSS code
 is totally removed.
\end{abstract}
Due to the Hesenberg uncertain principle,
quantum key distribution is different from classical cryptography in that 
an unknown quantum state is  in principle not known to Eve unless it is 
disturbed, rather 
than the conjectured difficulty of computing certain functions.
The first published protocol, proposed in 1984~\cite{BB}, 
is called BB84 after its inventors (C. H. Bennett and G. Bras\-sard.) For a 
history of the subject, one may see e.g. \cite{gisin}.
In this protocol, the participants (Alice and Bob)
wish to agree on a secret key about which no
eavesdropper (Eve) can obtain significant information.  
Alice sends each bit of
the secret key in one of a set of conjugate bases which 
Eve does not know, and this key is protected by the impossibility of
measuring the state of a quantum system simultaneously in 
two conjugate bases. Since then, studies on QKD are extensive.
In particular, the strict proof on the unconditional security have abstracted
much attentions. 
The original papers proposing quantum key distribution~\cite{BB} proved 
it secure against certain attacks,
including those feasible using
current experimental techniques.  However, for many years, 
it was not rigorously proven secure
against an adversary able to perform any physical
operation permitted by quantum mechanics.

 The first general although rather complex
proof of unconditional security was given by Mayers\cite{mayersqkd}, which was followed by a
number of other proofs\cite{others,others2}.    
 Building on the quantum privacy amplification idea of
Deutsch {\it et al.}\cite{deutsch}, Lo and
Chau\cite{qkd},   proposed a conceptually
simpler proof of security.  This protocol, although has
a drawback of requiring a quantum computer, opens the possibility
of finding simple proofs on a prepare and measure protocol.
 Later on,
Shor and Preskill \cite{shorpre}  unified the techniques in
 \cite{mayersqkd} and \cite{qkd} and provided a simple proof of
security of standard BB84.  (See also~\cite{squeezed} for a detailed
exposition of this proof.)  Shor-Preskill's proof is a reduction
from the purification scheme to the quantum error correction with CSS code
\cite{CSS} and finally to the BB84 scheme of decoding  the classical
CSS code with one way classical communication. Very recently, movivated for  higher bit error rate tolerance
and  higher  efficiency, Gottesman and Lo\cite{gl} 
studied the two way communication entanglement
purification protocol(2-EPP) and proposed  
a 4 state prepare and measure protocol with the highest bit error tolerance 
among all modified BB84 protocols so far. They also significantly increased
the previous bit error tolerance rate for the six state protocol. 
The tolerable bit error  rate for six state protocol is then further improved
by Chau\cite{chau}. A general theorem on the sufficient
condition to convert a 2-EPP protocol to a classical one is also given 
in\cite{gl}.
However, it has not been studied there on how to remove 
 the complexity of decoding the
CSS code in their prepare and measure protocol. 
So far in all those protocols based on CSS code, in the decoding stage,
Bob has to compare his string  with an exponentially large number of
all the strings in certain code space to find out the one with the shortest
distance with his string. 
The complexity of such a comparison can be huge without any preshared secrect
string.
For example, if we try to distill a final key of 300 bits, the complexity
will be far beyond the power of any exisiting classical computer. 
Studies towards the removal of the decoding complexity 
are rare. To the best of our  knowledge 
the only report on this topic is given by
H.K. Lo\cite{pad}. However,  Lo's scheme requires that Alice and Bob
have a pre-shared secrect string.
Here we take a further study on the 2-EPP QKD\cite{gl} 
and we find that besides 
the advantage of a higher bit error tolerance as reported in\cite{gl}, the 2-EPP protocol
has another advantage, it can be used to  remove the huge complexity in decoding
the classical CSS code.
We will construct  a specific prepare and measure protocol without the decoding complexity.
That means, in our protocol, even a large $classical$  computer is unnecessary.  
Before going into details of our protocol, we first make some mathematical
notations and some background presumptions for the quantum key distribution.

We will use two level quantum states as our qubits. For example, spin half
particles or linearly polarized photons. A quantum state can be prepared or measured in different basis. We define the spin up, down or
polarization of horizontal, vertical as the $Z$ basis, i.e., the basis of
$\{|0\rangle,|1\rangle\}$. We define the spin right, left or
polarization of $\pi/4,3\pi/4$  as the $X$ basis, i.e., the basis of
$\{|+\rangle,|-\rangle\}$. These basis are related by
$|+\rangle=\frac{1}{\sqrt 2}(|0\rangle + |1\rangle)$ and
$|-\rangle=\frac{1}{\sqrt 2}(|0\rangle - |1\rangle)$.\\
There are  four maximally entangled states(Bell basis)
\[
\Psi^\pm = \frac{1}{\sqrt{2}}( |01\rangle \pm |10\rangle),
\quad
\Phi^\pm = \frac{1}{\sqrt{2}}( |00\rangle \pm |11\rangle),
\]
which form an orthonormal basis for the quantum state space of two qubits.\\
 There are  three Pauli matrices:
\[
\sigma_x = \left(\begin{array}{rr}0 & 1 \\ 1 & 0 \end{array} \right),
\quad
\sigma_y = \left(\begin{array}{rr}0 & -i \\ i & 0 \end{array} \right),
\quad
\sigma_z = \left(\begin{array}{rr}1 & 0 \\ 0 & -1 \end{array} \right).
\]
The matrix $\sigma_x$ applies a bit flip error to a qubit, while
$\sigma_z$ applies a phase flip error.
We denote the Pauli matrix $\sigma_a$ acting 
on the $l$'th bit of the CSS code
by $\sigma_{a(l)}$ for $a \in \{x,y,z\}$.
For a binary vector~$\gamma$, we let 
\[
\sigma_a^{[\gamma]} = 
\sigma_{a(1)}^{\gamma_1}
\otimes
\sigma_{a(2)}^{\gamma_2}
\otimes
\sigma_{a(3)}^{\gamma_3} 
\otimes\ldots\otimes
\sigma_{a(n)}^{\gamma_n}
\]
where $\sigma_a^0$ is the identity matrix and $\gamma_i$ is the $i$'th 
bit of~$\gamma$.
The matrices $\sigma_x^{[s]}$ ($\sigma_z^{[s]}$) have 
all eigenvalues $\pm 1$.\\
We also need a short review the properties of CSS code\cite{CSS}.
Here we directly borrow the review materials given in ref.\cite{shorpre}.
Quantum error-correcting codes are subspaces of the Hilbert 
space ${\bf C}^{2^n}$ which are protected from errors in a small 
number of these qubits, so that any such error can be measured and 
subsequently corrected without disturbing the encoded state.  
A quantum CSS code $Q$ on $n$ qubits comes from two binary codes on $n$ 
bits, $C_1$ and $C_2$, one contained in the other: 
\[
\{0\} \subset {C_2} \subset {C_1} \subset {\bf F}_2^{n},
\]
where ${\bf F}_2^{n}$ is the binary vector space on $n$ bits \cite{CSS}.

A set of basis states (which we call {\em codewords}) for the CSS code 
subspace can be obtained from vectors $v \in C_1$ as follows:
\begin{equation}
{v}\, \longrightarrow\, \frac{1}{|C_2|^{1/2}} \sum_{w \in C_2} |v+w\rangle.
\label{codewords}
\end{equation}
If $v_1 - v_2 \in C_2$, then the codewords corresponding
to $v_1$ and $v_2$ are the same.  Hence these codewords 
correspond to cosets of $C_2$ in $C_1$, and this code protects
a Hilbert space of dimension $2^{\dim C_1 - \dim C_2}$.
Moreover, there is  a class of quantum error correcting codes equivalent
to $Q$, and parameterized by two $n$-bit binary vectors
$x$ and $z$.  Suppose that $Q$ is determined as above by $C_1$ and $C_2$.
Then $Q_{x,z}$ has basis vectors indexed by cosets of $C_2$ in $C_1$,
and for $v \in C_1$, the corresponding codeword is
\begin{equation}
{v} \,\longrightarrow\,|\xi_{v,z,x}\rangle= \frac{1}{|C_2|^{1/2}}
\sum_{w \in C_2} (-1)^{z\cdot w} |x+v+w\rangle.
\label{codewords2}
\end{equation}
We now make some presumptions. Without any loss of generality,
we assume a Pauli channel between Alice and Bob. All Eve's action can be regarded as (part of)  channel noise. A pauli channel is a channel acts independently
on each qubit by the Pauli matrices with classical probability. We shall
only consider two independent errors which are $\sigma_x$ error(bit flip error) and $\sigma_z$(phase flip)
error. All $\sigma_y$ error can be regarded as the joint error of $\sigma_x$
and $\sigma_z$.
Although the channel is noisy, we assume all qubits stored by Alice are never
corrupted. Moreover, we assume the classical communication between Alice and Bob is noiseless.

We start from  recalling Lo and Chau's protocol\cite{qkd} based on the entanglement
purification\cite{BDSW}. Suppose initially Alice and Bob share some impure EPR pairs. They randomly select a subset of them to check the bit flip
error rate and the phase flip error rate. They then distill a small number
of almost perfect EPR pairs from the remained pairs. They obtain the final key by
measuring them in each side in $Z$ basis. Note that here the only thing that is important
is to distill some almost perfect EPR pairs, it does not matter on how Alice
prepares the initial state. Actually, the random Hadamard transform on the
code qubits in Shor-Preskill protocol is uncessary. Note that every
qubit in transmission has the same density operator. 
In intercepting the qubits from Alice, Eve has neither  classical
information nor quantum information to distinguish which ones are check bits
and which ones are code bits. Eve cannot treat them differently. 
Therefore the bit flip error rate and the phase flip error rate in the code
bits must be close to that in the check bits, given a large number of
check bits and code bits. All these properties are not unchanged no matter
whether Alice takes random Hadamard transformation to the code bits which
are sent to Bob. 
We therefore have the following modified Lo-Chau-Shor-Preskill scheme:\\
{\bf Protocol 1: Modified Lo-Chau-Shor-Preskill protocol}
\begin{itemize}
\setlength{\itemsep}{-\parskip}
\item [\bf 1:] Alice creates $2n$ EPR pairs in 
        the state $(\Phi^+)^{\otimes n}$.
\item [\bf 2:] Alice sends the second half of each EPR pair to Bob.
\item [\bf 3:] Bob receives the qubits and publicly announces this fact.
\item [\bf 4:] Alice selects $n$ of the $2n$ encoded EPR pairs to serve as
	check bits to test for Eve's interference. In using the check bits,
she just randomly chooses the Z or X basis to measure and tells Bob does the same measurement to his halves on the same basis.  They compare the measurement
result on each check qubits.
If too many of these measurements outcomes disagree, they abort the
        protocol.  
\item [\bf 5:] Alice and Bob make the measurements on their code qubits
        of $\sigma_z^{[r]}$ for each row $r \in H_1$ and $\sigma_x^{[r]}$ for
        each row $r \in H_2$.  Alice and Bob share the results, compute
        the syndromes for bit and phase flips, and then transform their 
        state so as to obtain some nearly perfect EPR pairs.  
\item [\bf 6:] Alice and Bob measure the EPR pairs in the 
        $|0\rangle$, $|1\rangle$ basis to obtain a shared secret key.
\end{itemize}
Different from that in\cite{shorpre}, here Alice does not take any  random Hadamard transform to the code qubits sent to Bob.
Actually, step 5 can be replaced by a two way communication purification
scheme satisfying certain restrictions\cite{gl}. 
Morever, it can be divided into two steps, i.e.,  correcting
$all$ the bit flip error first and then  correcting the phase flip errors. 
Now we show how to correct $all$ bit flip errors. 
We shall call a purification protocol using the above two steps as the
{\it extremely unsymmetric} protocol in comparison with the normal protocols 
correcting bit flips and phase flips alternatively. 
This includes two stages:\\
{\bf 1. Crude bit flip error correction correction:}
Sharing a large number( say, $n$) of imperfect EPR pairs with known
upper
bound of bit flip  error rate,
Alice and Bob may just randomly pick out two pairs( $j$ and $k$) and compare the 
 parity. More specifically, they  take a controlled-not 
 operation $U_c$ on  
each side(they use qubit $j$ as the control qubit and qubit $k$ as the target
qubit). They each  meassure the target bit,  $k$ in $Z$ basis and compare the value(see figure 1).  $U_c$ here is defined as
\begin{eqnarray}
U_c|x_j,x_k\rangle = |x_j, x_j\oplus x_k\rangle
\end{eqnarray}
where $|x_j,x_k\rangle$ is any possible quantum state for qubits $j$ and $k$, expressed
in  $Z$ basis. 
If the
values on each side are same, they drop the target qubit $k$ and keep the control 
qubit in a new set $d_1$. If the values are different, they drop both qubits. 
They then randomly pick out another two pairs from the remained 
$n-2$ pairs and check the parity again by the
controlled-not gate and measurement on the target qubits in Z basis 
in each side.
They can repeatedly 
do so until they have picked out all $n$ imperfect pairs.
If the original bit error rate for the $n$ imperfect pairs is $\epsilon_b$,
the new bit error rate in the set $d_1$ is now reduced to a little bit higher
than $\epsilon_b^2$. They can take the same parity check action to the qubits 
in the new set $d_1$. They can take the similar action iteratively 
until  they believe that the bit flip error rate in the remained qubits
have been decreased to a  very small value, e.g., $10^{-3}$ (or $10^{-4})$. They then divide
their qubits into a number of subset $\{S_i\}$, e.g., each subset includes 100 ( or 1000 ) qubits. There must be some subsets where the bit flip errors have been $all$ corrected. Now they have to find out those subsets whose bit errors have been all corrected.
\\{\bf 2. Verification of zero bit flip error:} 
The task now is to find out which subsets 
have been corrected perfectly on bit flip errors. 
We can use the verification scheme by asking the fair questions
used in\cite{qkd}. 
Lets consider an arbitrary subset $S_i$. Suppose there are $n_s$
qubits in this subset. Zero bit flip error on this
subset  means that, $if$ Alice and Bob $meassured$ each of them in $Z$ basis, 
they $would$
share a common string $s_i=s_{iA}=s_{iB}$. Here $s_{iA}$ and $s_{iB}$ are 
the strings for  bit values at Alice's side and Bob's side, respectively.
$Suppose$ they each $had$ meassured their qubits of $S_i$.  To verify
$s_{iA}=s_{iB}$ is equivalent to verify that $s_{i0}=s_{iA} \oplus
 \bar{s}_{iB}=r_0$, where $r_0$
is a string with all elements $1$. i.e. $r_0=111\cdots1$ and 
$\bar s_{iB}=r_0 \oplus s_{iB}= $.  
To verify a classical tring $s_{i0}=r_0$, Alice may generate $m$ random
strings $\{R_i\}$ in the same length with $s_{i0}$, where each bit value in
the random strings  $\{R_j\}$ are determined by a coin tossing. One can calculate the value $ s_{i0}\cdot R_j$. If all $R_i$ satisfies
\begin{eqnarray}
s_{i0}\cdot R_j(mod 2)= P(R_j)
\end{eqnarray}
 $s_{i0}$ must be identical to $r_0$ with a probability $1-2^{-m}$. Here
$P(R_i)$ is the parity of string $R_i$. 
In our EPP protocol, we have to verify that there is no bit flip error
for the $n_s$ pairs in the subset $\{S_i\}$. 
It is easy to see that
\begin{eqnarray}
s_{i0}\cdot R_j(mod 2)=\left(s_{iA}\cdot R_j\oplus s_{iB}\cdot R_j\oplus r_0\cdot R_j\right)(mod 2).
\end{eqnarray}
Therefore the condition that $s_{i0}\cdot R_j(mod 2)=P(R_j)$ is equivalent to
\begin{eqnarray}
s_{iA}\cdot R_j(mod 2)= s_{iB}\cdot R_j(mod 2),
\end{eqnarray}
where we have used the fact that $ r_0\cdot R_j(mod 2)=P(R_j)$.
To verify the above formula, Alice and Bob actually need not measure each of
the
qubits  in $Z$ basis. As we are showing now, 
they can first take the controlled
not operations in each side and gather the information of $s_{iA}\cdot R_j(mod 2)$ and  $s_{iB}\cdot R_j(mod 2)$ to one qubit in each side therefore the measurement is only done on one qubit in each side.  
Alice may first create $m$
classical random string $\{R_j\}$ and announce them. The length of $R_j$ are
$n_s,n_{s}-1\cdots n_{s}-m$ respectively. They first use the random
string $R_1$. Suppose all those bits in $R_1$ with bit value
1 are on the position $p_1,p_2\cdots p_k$(normally $k$ is around 
$n_s/2$), Alice and Bob each  do a controlled unitary transformation
$U'_c$ on qubits at  the position
 $p_1,p_2\cdots p_k$  in $S_j$. They use qubit $p_k$
in each side as the target qubit(see figure(2)). The unitary operator
$U'_c$ is defined by
\begin{eqnarray}
U'_c|x_{p_1},x_{p_2}\cdots x_{p_k}\rangle=|
x_{p_1},x_{p_2}\cdots x_{p{k-1}},x'_{p_k}\rangle.
\end{eqnarray} and
\begin{eqnarray}
x'_{p_k}=\sum_{j=1}^k x_{p_j}.
\end{eqnarray}
Here  $|x_{p_1},x_{p_2}\cdots x_{p_k}\rangle$ is a quantum state in
$Z\otimes Z\cdots Z$ basis.
Unitary transformation $U_c'$ replaces the state of $k$th qubit by the parity of all the
qubits of $p_1,p_2\cdots p_k$ in $Z$ basis and keep all other qubits unchenged.
 Alice and Bob then
measure the  qubit at the position $p_k$ in each side in $Z$ basis. 
The outcomes are just $s_{iA}\cdot R_j(mod 2)$ and  
$s_{iB}\cdot R_j(mod 2)$,
respectively.
If they are different,
they discard all qubits which are originally in  $S_i$. If they are 
 identical,
 they discard qubit $p_k$ in $S_i$  and change the  qubit index $l$ into
 $l-1$  for any $l>p_k$ in $S_i$. Now the qubit index is from 1 to $n_s-1$.
They
use random string $R_2$ to redo the similar operation as that with string 
$R_1$. They take the operations repeatedly until they have exausted
all $R_j$ (or discard  all qubits which are originally in $S_j$ whenever they
find the values of the target bits in the two sides are different).
If the target bit values in two sides  are always identical, they accept the 
remained $n_s-m$ qubits in subset $S_i$. Now the probability
of {\it no bit flip error} for the survived qubits in $S_i$ is 
$1-2^{-m}$.
Suppose after the crude bit flip error correction the bit flip error
is $\epsilon_b^c$  and $n_s \epsilon^c_b<<1$, the probability of discarding
 $S_i$ is a little bit larger than $n_s\epsilon^c_b(1-\epsilon^c_b)^{n_s-1}$
after the {\it verification} stage.
Note that after this bit flip error correction,
 the phase flip error for the remained
qubits is increased. We denote the new phase error rate by $\epsilon'_p$. 
Suppose before any error correction, the bit flip error rate is $\epsilon_b$
and the phase flip error rate is $\epsilon_p$ and the joint error($\sigma_y$ type error) rate is $\epsilon_{bp}$. The $prior$ probability for a qubit carrying a phase flip error but no bit flip error is $\epsilon_p-\epsilon_{bp}$. 
After all bit flip errors are corrected(i.e., allr qubits carrying only a bit flip error and
all qubits carrying both errors are removed), the post probability for a qubit carrying a  phase flip error is
$$
\epsilon'_p=\frac{\epsilon_p-\epsilon_{bp}}{1-\epsilon_b}
$$
Obviously, the worst case  $\epsilon_{bp}=0$ 
leads to the highest value of $\epsilon'_p$. Therefore the
upper bound for the new phase flip error rate after the bit flip error correction is 
\begin{eqnarray}
\epsilon'_p=\frac{\epsilon_p}{1-\epsilon_b}.
\end{eqnarray}
Note that once $all$ bit flip errors are corrected,
the bit flip error will not increase any more by the subsequent
phase flip error correction.  
Protocol 1 is now reduced to the following protocol
\\
{\bf Protocol 2: Extremely unsymmetric distillation protocol}
\begin{itemize}
\setlength{\itemsep}{-\parskip}
\item [\bf 1:] Alice creates $2n$ EPR pairs in 
        the state $(\Phi^+)^{\otimes n}$.
\item [\bf 2:] Alice sends the second half of each EPR pair to Bob.
\item [\bf 3:] Bob receives the qubits and publicly announces this fact.
\item [\bf 4:] Alice selects $n$ of the $2n$ encoded EPR pairs to serve as
	check bits to test for Eve's interference. In using the check bits,
she just randomly chooses the Z or X basis to measure and tells Bob does the same measurement to his halves on the same basis.  They compare the measurement
result on each check qubits. They find the  detected bit flip error rate is
$\epsilon_b$ and the phase flip error rate is $\epsilon_p$. 
If these values exceed certain threshold set in advance, they abort the
        protocol.  
\item [\bf 5:]Alice and Bob first use the 
crude bit flip error correction to reduce the bit flip error rate to
$\epsilon^c_b$ and
then  divide the remained qubits into $q$ subsets, suppose there are
$n_s$ qubits in each subset. They then use the {\it
veryfication of zero bit flip } scheme as described above to distill
a number of qubits where  bit flip error is zero. Suppose $g$ subsets have passed
the verification, Alice and Bob is now sharing $g(n_s-m)$ qubits whose bit flip
error rate is strictly 0 with a probability of $1-g\cdot 2^{-m}$ and phase
flip error rate is $\epsilon'_p$.
\item [\bf 6:] Alice and Bob make the measurements on their code qubits
        of $\sigma_z^{[r]}$ for each row $r \in H_1$ and $\sigma_x^{[r]}$ for
        each row $r \in H_2$.  Alice and Bob share the results, compute
        the syndromes for bit and phase flips, and then transform their 
        state so as to obtain $m$ nearly perfect EPR pairs.   
\item [\bf 7:] Alice and Bob measure the EPR pairs in the 
        $|0\rangle$, $|1\rangle$ basis to obtain a shared secret key.
\end{itemize}
Protocol 2 is a CSS like protocol\cite{gl}. In particular, all operations
including the controlled unitary transformations and measurements
in step 5 only are done only in $Z$ basis therefore the protocol satisfies
the main theorem in\cite{gl}. Consequently, this protocol can be converted
to the prepare and measure protocol, i.e. BB84 protocol.

In particular, using the arguments in Ref.\cite{shorpre}, step 6 and 7
in protocol 2 can be reduced to the encoding
and decoding of quantum CSS code and can be further reduced to 
a prepare and measure protocol followed by decoding a  CSS code with one way classical communication. 
Step 6 and 7 are equivalent to the case that Alice starts with $g(n_s-m)$ perfect
EPR pairs and send the second halves  to Bob through an unsymmetric noisy
channel causing no bit flip error and a phase flip error rate
bounded by $\epsilon'_p$. After
Bob received the  qubits from Alice they meassure the syndromes and then distill a small number of perfect EPR pairs. As argued in Ref.\cite{shorpre}, such a
process is equivalent to the process that Alice meassures each of her qubits
in $Z$ basis at any time and then Alice and Bob obtain the final key
by  decoding
a classical CSS code with one way classical communication. 
Specifically, step 6 and step 7 are equivalent to the following steps:\\
6': Alice measures all her qubits in Z basis and obtain a $g(n_s-m)-$bit
 state $|x\rangle$.
She randomly pick out a binary vector $v$ in code space $C_1$. She sends the 
binary classical string $x+v$ to Bob. \\
7': Bob measures his qubits in $Z$ basis and obtain $|x\rangle$ which is exactly
identical to Alice's measurement outcome with a probability $1-g\cdot 2^{-m}$. With
such a high probability that his state is identical to Alice's, he simply 
always assumes that there is no deviation between his measurement result 
and Alice's result. Using the information $x+v$ from Alice, he has a new string $v$ in code space $C_1$. \\
8: Alice and Bob use the coset of $v+C_2$ as their final key.     
\\Prior to step 6', all bit flip error had been removed, 
we only require our CSS code
used there to correct $\epsilon'_p$ phase flip error and 0 bit flip error.
Therefore we can safely set $dim(C_1)=g(n_s-m)$ in the CSS code.\\
Furthermore, step 5 is now followed immediately by Alice's 
measurement in Z basis
to all of her qubits. Since all operations in step 5 are in $Z$ basis, we can change the order of all these operations. In particular, Alice may choose to
meassure all of her qubits in the begining of step 5. 
This is equivalent to take measurement in $Z$ basis to all her code qubits in the begining of the whole protocol. If she does so, The controlled unitary transformation and all the parity checks can be done classically as the following:

{\bf 1. Classical crude bit flip error correction correction:}
Bob measures all his code qubits in $Z$ basis and obtain a classical string $s$. 
Alice and Bob  randomly pick out two  bits($x_j,x_{k}$) in the string and compare  parity. If the
values on each side are same, they drop $x_{k}$ and keep 
$x_j$ in a new set $d_1$. If the values are different, they drop both bits. 
They then randomly pick out another two bits from the remained 
$n-2$ bits in string $s$ and check the parity. If the parity is same, they drop one and place another one in set $d_1$. If the parity is different, they drop
both bits. 
They can repeatedly 
do so until they have picked out all bits in string $s$.
If the original bit error rate in string $s$ is $\epsilon_b$,
the new bit error rate in the set $d_1$ is now reduced to a little bit higher
than $\epsilon_b^2$. They can take the same parity check action to the bits in the new set $d_1$ and place all distilled bits in another set $d_2$. They can take the similar action iteratively 
until  they believe that the bit flip error rate in the remained bits
have been decreased to a  very small value, e.g., $10^{-3}$( or $10^{-4}$). 
They then divide
their bits into a number of substrings $\{S_i\}$, e.g., each substring
 includes 100 bits (or 1000 bits). There must be some substrings where the bit flip errors have been $all$ corrected. Now they start to find out those 
substrings whose bit 
flip errors have been all corrected.
\\{\bf 2. Classical verification of zero bit flip error:} 
Lets consider substring $S_i$. Suppose there are $n_s$
bits in this substring. Suppose $s_{iA}$ and $s_{iB}$ are 
the classical strings  at Alice's side and Bob's side, respectively.
 Alice  creates $m$
classical random string $\{R_j\}$ and announces them. The length of $R_1,R_2\cdots R_m$ are
$n_s,n_{s}-1\cdots n_{s}-m$, respectively. They first use the random
string $R_1$.  Suppose the last non-zero bit in $R_1$
is at position $p_k$. They each calculate the value $s_{iA}\cdot R1(mod2)$ and
$s_{iB}\cdot R_1(mod2)$ respectively. If they get the same result,
 they discard bit $p_k$ in $S_i$ and keep all the others and change the 
bit index
of $l$ into $l-1$ for any $l>p_k$ in $S_i$. Now there are only $n_s-1$ bits
remained in string $S_i$.
If they get a different result, they discard the whole $S_i$. 
They take the operation repeatedly until they exaust
all $R_j$ (or discard   $S_i$ once the have got the different value).
If $s_{iA}\cdot R_{j}(mod2)=s_{iB}\cdot R_j(mod2)$ 
for all $R_j$, they accept the 
remaining $n_s-m$ bits in substring $S_i$. Now the probability of
{\it no bit flip error}  is 
$1-2^{-m}$ for the survived bits in $S_i$. 
Suppose after the classical crude bit flip error correction the bit flip error
rate is $\epsilon^c_b$  and $n_s \epsilon^c_b<<1$, the probability of discarding
 $S_i$ is a little bit larger than $n_s\epsilon^c_b(1-\epsilon^c_b)^{n_s-1}$
after the {\it classical verification} stage. There must be a significant
number of substrings that can pass the {\it classical verification} check provided
the total bit flip error rate is rather small after the {\it classical crude
bit flip error correction}.\\
Again, as it was argued in\cite{shorpre},  Alice may also chooses to meassure
all her check qubits on $Z$ basis in the begining of the protocol. If she does so,
protocol 2 is equivalent to BB84 protocol, with a post selection on which
ones are check qubits, which ones are code qubits and which ones are qubits 
measured in wrong basis which should be discarded immediately. Therefor we have the following final
prepare and measure protocol:
\centerline{{\bf Protocol 3: Simplified BB84}}
\begin{itemize}
\setlength{\itemsep}{-\parskip}

\item [\bf 1:]Alice generates a classical set $W=\{1,2,3,4\}$.
She randomly picks out one value from this set. If she gets 2,3 or 4,
she prepares  a state in basis $\{|0\rangle, |1\rangle\}$.
 If she gets 1,
she prepares  a state in basis $\{|+\rangle, |-\rangle\}$.
 Alice creates $(4+\delta)n$ states in this way.  
\item [\bf 2:] Alice sends the resulting qubits to Bob.

\item [\bf 3:] Bob receives the $(4+\delta)n$ qubits, measuring each
of them in a basis randomly chosen from $X,Z$ by a coin tossing.

\item [\bf 4:] Alice announces the basis information for each qubit.

\item [\bf 5:] Bob discards any results where he measured in a different
	basis than Alice prepared.  With high probability, there are at
	least $2n$ bits left (if not, abort the protocol).  Bob chooses
all those remained qubits measured in the $|+\rangle$,$|-\rangle$ 
        basis and randomly chooses the same number of 
qubits  measured
in $Z$ basis as the check bits. They discard a few qubits and use the rest
$n$ qubits as the code bits.

\item [\bf 6:] Alice and Bob announce the values of their check bits.
	If too few of these values agree, they abort the protocol.
They find the bit flip error rate and the phase flip error
rate on the checked bits are $\epsilon_b$ and $\epsilon_p$ respectively.

\item [\bf 7:] They use the classical crude bit flip error correction scheme
and the classical verification of zero bit flip error scheme to distill
$g(n_s-m)$ bits. There are strictly no bit flip error for these $g(n_s-m)$ bits
with a probability $1-g\cdot 2^{-m}$. The new phase flip error is  bounded by
$\epsilon_1=\epsilon'_p+\eta$ with a probability larger than
$1-\exp(-\frac{1}{4}\eta^2 n/(\epsilon_p-\epsilon_p^2))$.
\item [\bf 8:] Alice announces $x+v$, where $x$ is a $g(n_s-m)-$bit 
binary string
        consisting of the measurement outcome for the remaining  bits, and $v$ is a random 
        binary string of  $g(n_s-m)$ bits.

\item [\bf 9:] Bob subtracts $x+v$ from his code qubits, $x$, and
        obtains $v$.  

\item [\bf 10:] Alice and Bob use the coset of $v + C_2$ as the final key.

\end{itemize}
 
This is a modified BB84 protocol. Here Bob measures $all$ the code bits in 
$Z$ basis instead of in the random basis $Z$ or $X$ used in the original
BB84.
In using the above protocol, the suceeding probability is larger than
$(1- 2^{-m}g)[1-\exp(-\frac{1}{4}\eta^2n/(\epsilon_p-\epsilon_p^2))]$.
In making the crude error correction to bit flip error, the number of qubits
in set $d_1$ will be less than $n/2$, that in $d_2$ will be less than 
$n/4$. The method of crude distillation plus verification is not necessarily
the most efficient one. It should be interesting to find out the most efficient scheme to make the quantum key distribution without classical complexity.

In summary, we have spotted that the random Hadamard transformation on the code
qubits sent to Bob is unnecessary in Alice's state preparation in an EPP protocol for quantum key distribution. Based on this fact, we have taken a further
study on the 2-EPP QKD protocol and we have constructed a prepare and measure 
QKD  protocol where the  bit filp correction  and the phase flip error correction(privacy amplification) is totally decoupled therefore the complexity of
CSS code decoding is totally avoided.

{\bf Acknowledgement:} I thank Prof Imai H for support. 

\begin{figure}
\epsffile{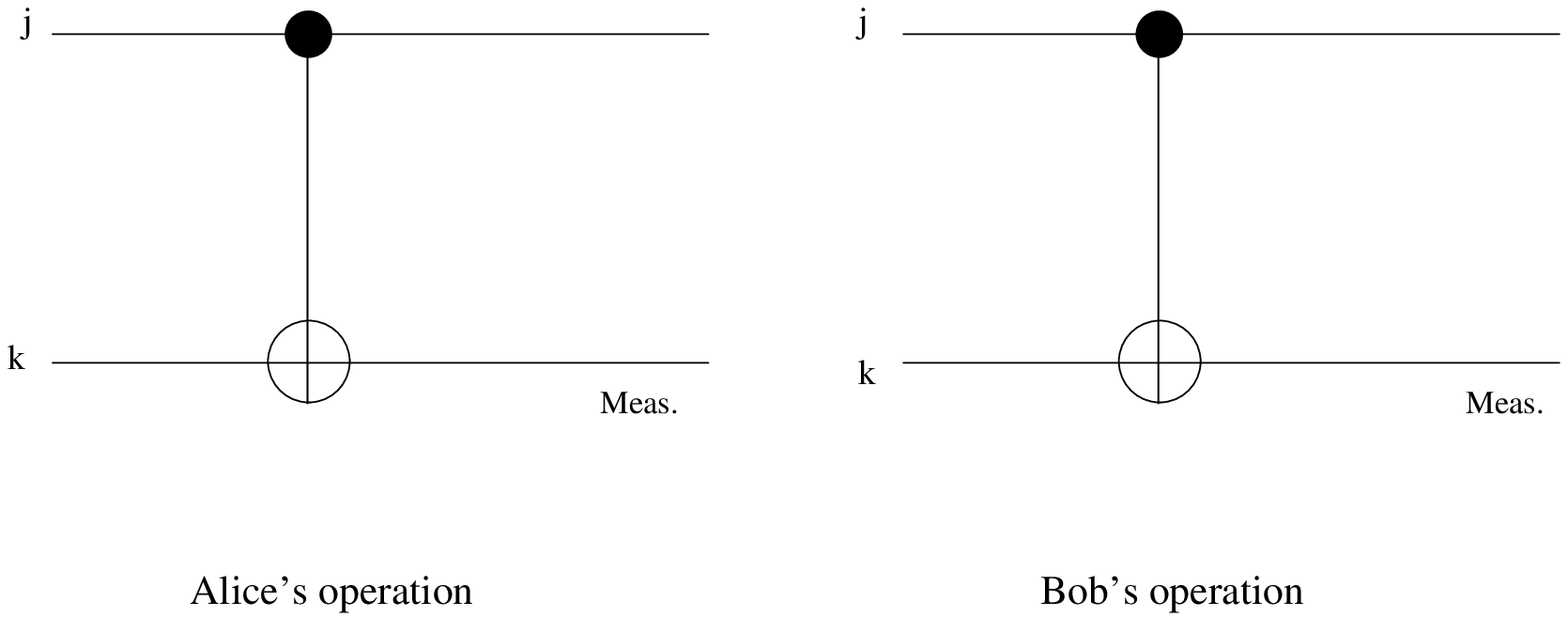}
\caption{ Controlled not operation used for the crude bit flip error correction. The horizontal lines marked by j and k are qubit $j$ and $k$ respectively.
 Alice and Bob  compare the measurement outcomes of the target qubit $k$.} 
\end{figure}
\begin{figure}
\epsffile{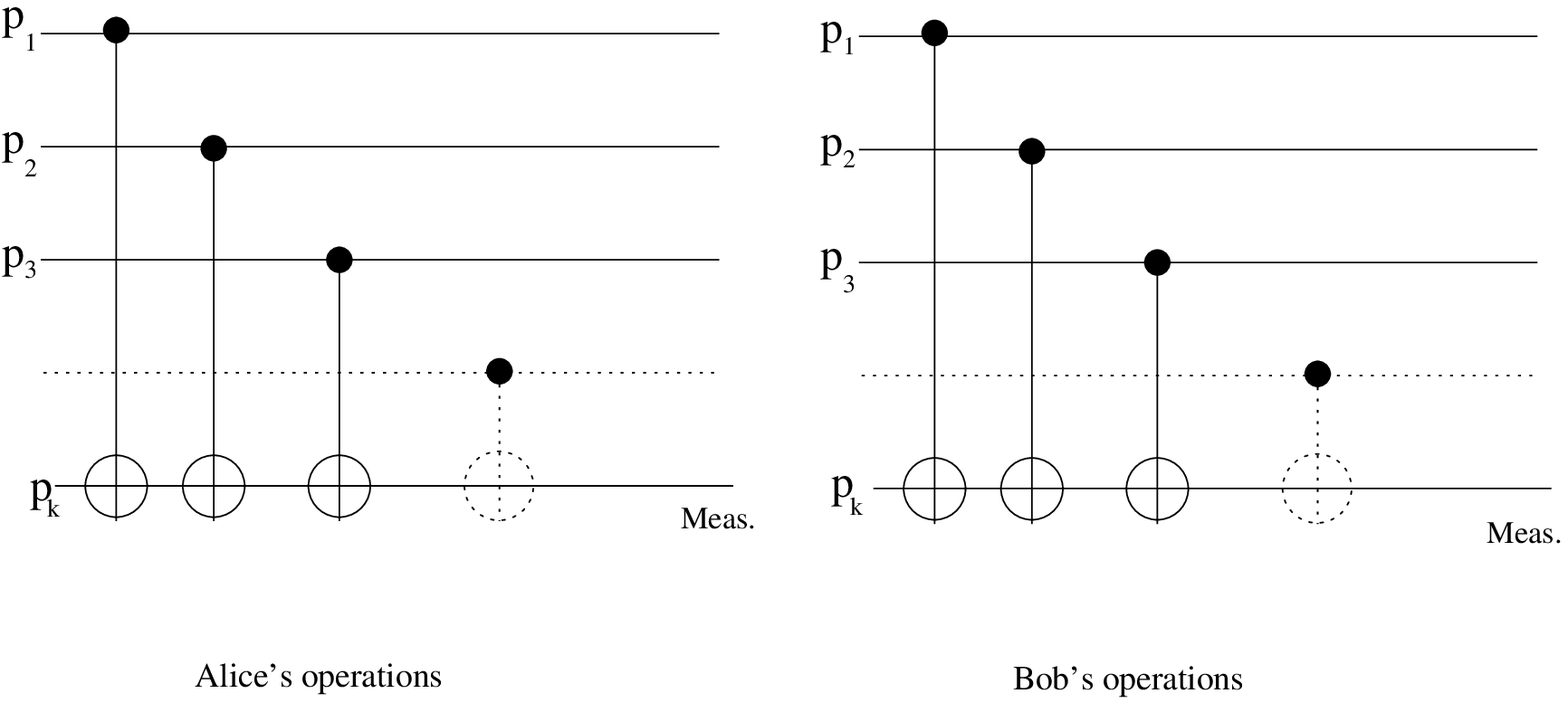}
\caption{ Controlled unitary operation used for the verification of zero
 bit flip error. The horizontal lines marked by $p_i$s are qubits at 
position $p_i$s in set $S_j$. Alice and Bob compare the measurement
 outcomes of the target qubit $p_k$. } 
\end{figure}

\begin{thebibliography}{99}
\bibitem{BB}
C. H. Bennett and G. Brassard, 
``Quantum cryptography: Public-key distribution and coin tossing,'' 
in {\em Proceedings of IEEE International Conference on Computers, 
Systems and Signal Processing, Bangalore, India, 1984},  (IEEE Press,
1984), pp. 175--179;
C.H. Bennett and G. Brassard,
``Quantum public key distribution,''
IBM Technical Disclosure Bulletin {\bf 28}, 3153--3163 (1985).
\bibitem{gisin} N. Gisin, G. Ribordy, W. Tittel, and H. Zbinden,
``Quantum Cryptography,'', Reviews of Modern Physics, vol. 74, pp. 145-195.
Also [Online]
Available: http://xxx.lanl.gov/abs/quant-ph/0101098.
\bibitem{mayersqkd} D. Mayers, ``Unconditional security in
Quantum Cryptography,'' Journal of ACM, vol. 48, Issue 3, p. 351-406. 
Also [Online] Available:
http://xxx.lanl.gov/abs/quant-ph/9802025.
\bibitem{others}
E. Biham, M. Boyer, P. O. Boykin, T. Mor, and V. Roychowdhury,
``A proof of the security of quantum key distribution,''
in {\it Proceedings of
the Thirty-Second Annual ACM Symposium on Theory of Computing} (STOC)
(ACM Press, New York, 2000), p. 715.
\bibitem{others2}
M. Ben-Or, Unpublished.
\bibitem{deutsch} D.~Deutsch, A.~Ekert, R.~Jozsa, C.~Macchiavello,
S.~Popescu, and A.~Sanpera, ``Quantum privacy amplification and the
security of quantum cryptography over noisy channels,''
{\it Phys.~Rev.~Lett.}, vol.  77, p. 2818, 1996.
Also, [Online] Available:
http://xxx.lanl.gov/abs/quant-ph/9604039.  Erratum
Phys.~Rev.~Lett. {\bf 80}, 2022 (1998).
\bibitem{qkd} H.-K.~Lo and H.~F.~Chau, ``Unconditional security of
quantum key distribution over arbitrarily long distances,'' Science
\bibitem{shorpre} P. W. Shor and J. Preskill, ``Simple proof of
security of the BB84 quantum key distribution protocol,''
{\it Phys. Rev. Lett.}, vol. 85, p. 441, 2000. Also,
[Online] Available: http://xxx.lanl.gov/abs/quant-ph/0003004.
\bibitem{squeezed} D. Gottesman and J. Preskill, ``Secure
quantum key distribution using squeezed states,''
{\it Phys. Rev.}, vol. A63, p. 22309, 2001.
\bibitem{CSS}
A. R. Calderbank and P. Shor, 
``Good quantum error correcting codes exist,''
Phys. Rev. A {\bf 54}, 1098--1105 (1996),
{\em \mbox{arXive} e-print} quant-ph/9512032; 
A. M. Steane, 
``Multiple particle interference and error correction,'' 
Proc. R. Soc. London A
{\bf 452}, 2551--2577 (1996),
{\em \mbox{arXive} e-print} quant-ph/9601029.
\bibitem{gl} D. Gottesman and H.-K. Lo, quant-ph/0105121, ``Proof of
security of quantum key distribution with two-way classical communication''.
\bibitem{chau} H. F. Chau, in quant-ph/0206050, ``Practical scheme to share a secret key
through up to 27.6
\bibitem{pad}H. K. Lo, quant-ph/0201030, ``Method for decoupling
error correction from privacy amplication''.
\bibitem{BDSW} C. H. Bennett, D. P. DiVincenzo, J. A. Smolin,
and W. K. Wootters, ``Mixed state entanglement and quantum error
correction,'' {\it Phys. Rev.}, vol. A54, 3824, 1996.
\bibitem{bdsw}
C. H. Bennett, D. P. DiVincenzo, J. A. Smolin and W. K. Wootters,
``Mixed state entanglement and quantum error correction,''
Phys. Rev. A,  {\bf 54}, 3824--3851 (1996),
{\em \mbox{arXive} e-print} quant-ph/9604024.
\end{thebibliography}
\end{document}